\documentclass{ws-procs9x6}
\usepackage[english]{babel}
\usepackage{slashed}
\usepackage{amsmath}
\begin{document}

\vspace*{-2,5cm}
\title{\textbf{NUCLEON TO PION TRANSITION DISTRIBUTION AMPLITUDES IN A LIGHT-CONE QUARK MODEL}}
\author{\normalsize M.~\textsc{Pincetti}$^\ddagger$, B.~\textsc{Pasquini}, S.~\textsc{Boffi}}
\address{Dip. di Fisica Nucleare e Teorica, Universit\`a
degli Studi di Pavia\\ and INFN, Sezione di Pavia, Pavia, 27100,
Italy\\$^\ddagger$E-mail: manuel.pincetti@pv.infn.it}

\begin{abstract}
{We present a general representation for the nucleon distribution amplitudes and for the nucleon to pion transition distribution amplitudes in terms of light-cone wave functions. We apply our formalism to a light-cone constituent quark model giving some numerical results for both the classes of observables.}
\end{abstract}

 \bodymatter

\section{Introduction}

Until few years ago all matrix elements taken into account in
the description of exclusive reactions at high momentum transfer were those between
initial and final hadron states, giving rise to the so-called Generalized
Parton Distributions (GPDs)\cite{ReportGPDS}, or those involving an hadronic and a
vacuum state, describing the well-known Distribution Amplitudes
(DAs)\cite{Chernyak}. The generalization of such distributions to
the case where the initial and final states correspond to different
particles has recently been proposed. This new mathematical objects,
initially called skewed distribution amplitudes\cite{sda}, are now established
as Transition Distribution Amplitudes\cite{Pireetal} (TDAs), since they describe a
transition between two different hadronic states.

In this article, we perform a Fock-state decomposition of the nucleon
state with the aim to understand the structure of the
nucleon in term of its Light-Cone Wave Function (LCWF)
representation. First, we derive the general LCWF representation for
the nucleon DAs taking into account the lower Fock-state component
made up of three valence quarks only. Then, for the first time in the
literature we apply the LCWF formalism to the nucleon to pion TDAs
for which we are forced to consider at least the state composed of
five partons, \textit{i.e.} three valence quarks and a quark-antiquark pair
describing the mesonic cloud surrounding the nucleon.
The nucleon to pion TDAs are believed to be the most direct observables to test
the pion-cloud contribution to the nucleon structure. Their experimental study is ongoing (JLab) or is planned (GSI, CERN).

The outline of the paper is as follows. In section 2 we give a detailed description of the LCWF representation for the DAs in the valence sector and we give some numerical results for the DAs and their moments within a light-cone quark model, in comparison with other model calculations and lattice predictions. Then in section 3 in the framework of the convolution model of Ref.~\refcite{Barbara-MesonCloud} we present the representation of the nucleon to pion TDAs in terms of LCWFs giving also some numerical predictions for three of them.

\section{DAs in the LCWFs formalism}

\subsection{Definitions and properties}

In general the notion of DAs refers to hadron to vacuum matrix
elements of nonlocal operators built up of quark and gluon fields at
light-like separation. In this paper we deal with the three-quark
matrix element
\begin{equation}\label{General-Matrix-Element}
\langle0|\epsilon^{ijk}u^{i'}_\alpha(z_1n)[z_1;z_0]_{i'i}
u^{j'}_\beta(z_2n)[z_2;z_0]_{j'j}
d^{k'}_\gamma(z_3n)[z_3;z_0]_{k'k}|P(p_1,\lambda)\rangle,
\end{equation}
where $|P(p_1,\lambda)\rangle$ denotes the proton state with
momentum $p_1$ ($p_1^2=M^2$) and helicity $\lambda$. The Latin letters
$i, j$ and $k$ refer to colour, while the Greek letters $\alpha,
\beta$ and $\gamma$ stand for Dirac indices.
We define the light-cone vectors $p$
and $n$ ($p^2=n^2=0$) such that $2p\cdot n = 1$.
The most general
decomposition of the matrix element in
Eq.~(\ref{General-Matrix-Element}) involves 24 invariant functions\cite{Braunetal},
but to leading twist accuracy only three of them survive, \textit{i.e.}
\begin{eqnarray}\label{DA}
&&4\mathcal{F}\bigg(\langle0|\epsilon^{ijk}u^i_\alpha(z_1n)
u^j_\beta(z_2n)d^k_\gamma(z_3n)|P(p_1, \lambda)\rangle\bigg)  =
f_N\Big[V^p(\slashed{p}_1C)_{\alpha\beta}(\gamma^5N^+)_\gamma\nonumber\\ &&\;\;\;\;\;\;\;\;\;\;\;\;\;\;\;\;\;\;+
A^{p}(\slashed{p}_1\gamma^5C)_{\alpha\beta}(N^+)_\gamma +
T^{p}(\sigma_{p_1\mu}C)_{\alpha\beta}(\gamma^\mu\gamma^5N^+)_\gamma\Big],
\end{eqnarray}
where $\sigma^{\mu\nu} = 1/2[\gamma^\mu, \gamma^\nu]$,  $\sigma^{p_1\mu}$ is
a shorthand notation for $p_{1\nu}\sigma^{\nu\mu}$, $C$ is the charge
conjugation matrix, $N^+$ is the \lq\lq good'' or \lq\lq large''
component of the nucleon spinor $N$, $\mathcal{F}$ represents a Fourier transform\footnote{The symbol $\mathcal{F}$ is a shorthand notation for the Fourier transform $$(p\cdot
n)^3\int\frac{\prod_j\mathrm{d}z_j}{(2\pi)^3}\exp[ix_kz_kp\cdot n].$$}, and $f_N$ is the value of the nucleon wave function at the origin.
The three functions $V^p$, $A^p$ and $T^p$ depend on the variables $x_i$ ($0<x_i<1$, $\sum_ix_i=1$) which correspond to the longitudinal momentum fractions carried by the quarks inside the nucleon.
Because of the symmetry properties of the operator in Eq.~(\ref{DA}) it is easy to see that the three invariant functions can be expressed in term of a single function $\Phi$. Indeed, labelling with 1, 2 or 3 the arguments of the DAs the following relations hold:
\begin{eqnarray}\label{Symmetry_Properties}
V(1,2,3) = V(2,1,3),\;\, A(1,2,3) = -A(2,1,3),\;\, T(1,2,3) = T(2,1,3),
\end{eqnarray}
and
\begin{eqnarray}\label{Symmetry_Properties2}
2T(1,2,3) = \Phi(1,3,2) + \Phi(2,3,1),\;\;
\Phi(1,2,3) = V(1,2,3) - A(1,2,3).
\end{eqnarray}
Defining with the symbol
$M_{\alpha\beta,\gamma}^{\uparrow/\downarrow}$ the matrix element in
the left-hand side of Eq.~(\ref{DA}), we can express the
three DAs as\footnote{In principle one has to face with 128
matrix elements (4 $\times$ 4 $\times$ 4 $\times$ 2) because there
are three Dirac indices and two helicity states, but for symmetry
reasons only three of them are independent.}
\begin{equation}\label{V}
V^{p} = \frac{1}{f_N}\frac{1}{\sqrt[4]{2}}(p_1^+)^{-\frac{3}{2}}\Big(M_{12,1}^\uparrow + M_{21,1}^\uparrow\Big),
\end{equation}
\begin{equation}\label{A}
A^{p} = \frac{1}{f_N}\frac{1}{\sqrt[4]{2}}(p_1^+)^{-\frac{3}{2}}\Big(M_{21,1}^\uparrow - M_{12,1}^\uparrow\Big),
\end{equation}
\begin{equation}\label{T}
T^{p} = -\frac{1}{f_N}\frac{1}{\sqrt[4]{2}}(p_1^+)^{-\frac{3}{2}}M_{11,2}^\uparrow.
\end{equation}
Thus to calculate the leading twist DAs we must give an expression for the $M_{\alpha\beta,\gamma}^{\uparrow/\downarrow}$ matrix elements.

\vspace{-3mm}
\subsection{LCWFs representation}

The representation in terms of LCWFs has been applyed to a large range of observables like form factors\cite{Drell}, the transversity distribution\cite{transversity}, GPDs\cite{overlaprepresentation} and transverse momentum dependent distributions\cite{TMDs} as a uselful formalism to disentangle the contribution of the different partonic configurations in the nucleon. Here we sketch the derivation of the LCWF representation for the DAs.
Starting from the left-hand side of Eq.~(\ref{DA}) and substituting the general Fourier expansion in momentum space of the free quark field one finds,
\begin{eqnarray}\label{x_DA}
&&\langle0|4(p_1^+)^3\epsilon^{ijk}\int\frac{\mathrm{d}k_1^+\mathrm{d}^2\mathbf{k}_{1\perp}}{16\pi^3k_1^+}
\int\frac{\mathrm{d}k_2^+\mathrm{d}^2\mathbf{k}_{2\perp}}{16\pi^3k_2^+}\int\frac{\mathrm{d}k_3^+\mathrm{d}^2\mathbf{k}_{3\perp}}{16\pi^3k_3^+}
\Theta(k_1^+)\Theta(k_2^+)\Theta(k_3^+)\nonumber\\\nonumber&&\times\sum_{\lambda_1, \lambda_2, \lambda_3}
b^i_{1}(\tilde{k}_1, \lambda_1)b^j_{2}(\tilde{k}_2, \lambda_2)b^k_{3}(\tilde{k}_3,
\lambda_3)u_{+\alpha}(k^+_1, \lambda_1)u_{+\beta}(k^+_2,
\lambda_2)u_{+\gamma}(k^+_3,
\lambda_3)\nonumber\\&&\times\delta\Big(x_1p_1^+ - k_1^+\Big)\delta\Big(x_2p_1^+ -
k_2^+\Big)\delta\Big(x_3p_1^+ -
k_3^+\Big)|P(p_1,s_1)\rangle,
\end{eqnarray} 
where $\tilde{k}_i \equiv (k^+_i, \mathbf{k}_{i\perp})$ is a shorthand notation for the plus and transverse parton momentum
components, $\lambda_i$ represents the parton helicity and $b$ is the
annihilator of the \lq\lq good'' component of the quark fields. 
Truncating the Fock expansion to the minimal configuration corresponding to three valence quarks, the proton state in Eq.~(\ref{x_DA}) can be written as
\begin{eqnarray}\label{proton_x_DA}
|P(p_1, s_1)\rangle &=& 
\sum_{\lambda_i, \tau_i, c_i}
\int\prod_{i={1}}^{3}
\frac{\mathrm{d}y_i}{\sqrt{y_i}}
\int
\frac{\prod_{i={1}}^{3}\mathrm{d}^2\boldsymbol{k}_{i\perp}}{[2(2\pi)^3]^2}
\delta\bigg(1 - \sum_{i=1}^{3}
y_i\bigg)\delta^{(2)}\bigg(\sum_{i={1}}^{3}\boldsymbol{k}_{i\perp}\bigg)
\nonumber\\
&&\times\tilde{\Psi}_{\lambda}^{[f]}(\{y_i,
\boldsymbol{k}_{i\perp}; \lambda_i, \tau_i,
c_i\})\prod_{i={1}}^{3}|y_ip_1^+, \boldsymbol{k}_{i\perp} +
y_i\mathbf{p}_{1\perp}, \lambda_i, \tau_i, c_i; q\rangle,\nonumber\\
\end{eqnarray}
where $\lambda_i, \tau_i$ and $c_i$ are spin, isospin and colour
variables of the quarks, respectively.
Then, after some algebra one obtains\footnote{The colour dependence in the LCWF $\tilde{\Psi}_{\lambda}^{[f]}$ has already been taken into account in the numerical prefactor.}
\begin{eqnarray}\label{finalMDA}
M_{\alpha\beta,\gamma}^{\uparrow/\downarrow}&=&-\frac{24}{\sqrt{x_1x_2x_3}}
\sum_{\lambda_1,\lambda_2,\lambda_3} u_{+\alpha}(x_1p_1^+,
\lambda_1)u_{+\beta}(x_2p_1^+, \lambda_2)u_{+\gamma}(x_3p_1^+,
\lambda_3)\nonumber\\&&\times\int\frac{\prod_{i=1}^3\mathrm{d}^2\boldsymbol{k}_{i\perp}}{[2(2\pi)^3]^2}\delta^{(2)}\bigg(\sum_{i={1}}^{3}\boldsymbol{k}_{i\perp}\bigg) \nonumber\\&&\times\tilde{\Psi}_{\lambda}^{[f]}(\{x_1,
\boldsymbol{k}_{1\perp}; \lambda_1, 1/2\}\{x_2,
\boldsymbol{k}_{2\perp}; \lambda_2, 1/2\}\{x_3,
\boldsymbol{k}_{3\perp}; \lambda_3, -1/2\}).\nonumber\\
\end{eqnarray}
The previous equation is a formal general representation for the matrix
element involved in the DAs definition.
To get the final representation for the matrix
element what is missing is an expression for the LCWFs. In this
work we follow the same procedure adopted in
Refs.~\refcite{Barbara-Linking}, \refcite{Barbara_Papers}. We refer
to these works for the details and here we give the final
results for the DA $\Phi$
\begin{eqnarray}\label{Vfinal}
 \Phi &=&
\frac{8\sqrt{2}}{f_N}\bigg[\frac{1}{M_0}\frac{\omega_1\omega_2\omega_3}{x_1x_2x_3}\bigg]^{\frac{1}{2}}
\int\frac{\prod_{i=1}^{3}\mathrm{d}^2\boldsymbol{k}_{i\perp}}{16\pi^3}\delta^{(2)}\bigg(\sum_{i=1}^{3}\boldsymbol{k}_{i\perp}\bigg)\psi(\boldsymbol{k}_{1},\boldsymbol{k}_2,\boldsymbol{k}_3)\nonumber\\&&\times
\frac{\bigg\{\Big[a_1\kappa_{2}^R\kappa_{3}^L\Big]
+ \frac{1}{2}\Big[a_1a_2a_3\Big]-
\frac{1}{2}\Big[\kappa_{1}^L\kappa_{2}^Ra_3\Big]
\bigg\}}{\sqrt{\Big[a_1^2 +
\boldsymbol{k}_{1\perp}^2\Big]\Big[a_2^2 +
\boldsymbol{k}_{2\perp}^2\Big]\Big[a_3^2 +
\boldsymbol{k}_{3\perp}^2\Big]}},
\end{eqnarray}
where $a_i=m + x_iM_0$ and $\kappa_i^{L,R} = \kappa_{ix} \mp \kappa_{iy}$, with $m$ being the quark mass and $M_0$ the eigenvalue solution of the free Hamiltonian. For the description of the momentum dependence part $\psi(\boldsymbol{k}_{1},\boldsymbol{k}_2,\boldsymbol{k}_3)$ of the LCWF we adopt the parameterization of  Ref.~\refcite{Schlumpf}.

\vspace{-3mm}
\subsection{Numerical Results}

In the literature there are many different papers trying to give a good
description of the nucleon
DAs~\cite{DaPhi,KS,Gari_Stef,Dzi,Schafer,Dzi&Fran,Stef_Berg,Bolz_Kroll}. In particular, in Fig.~1 the DA $\Phi$ from 
the data fit of Ref.~\refcite{Bolz_Kroll} is compared with our model calculation. Our
result is in good agreement with the data fit of Bolz and Kroll.
\begin{figure}[ht]
\centerline{
\epsfig{file=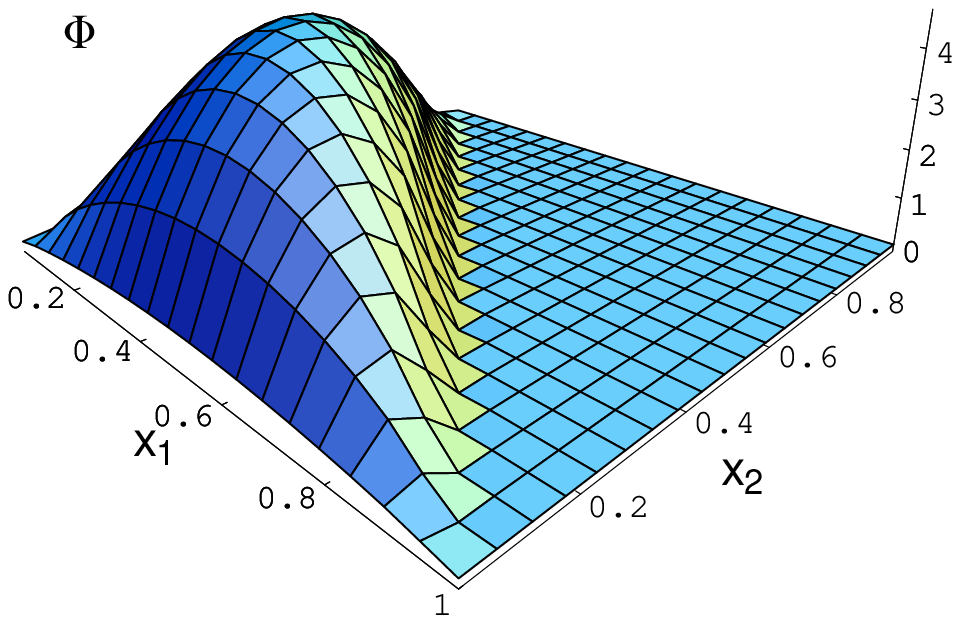,  width=13 pc} \hspace{1pc}
\epsfig{file=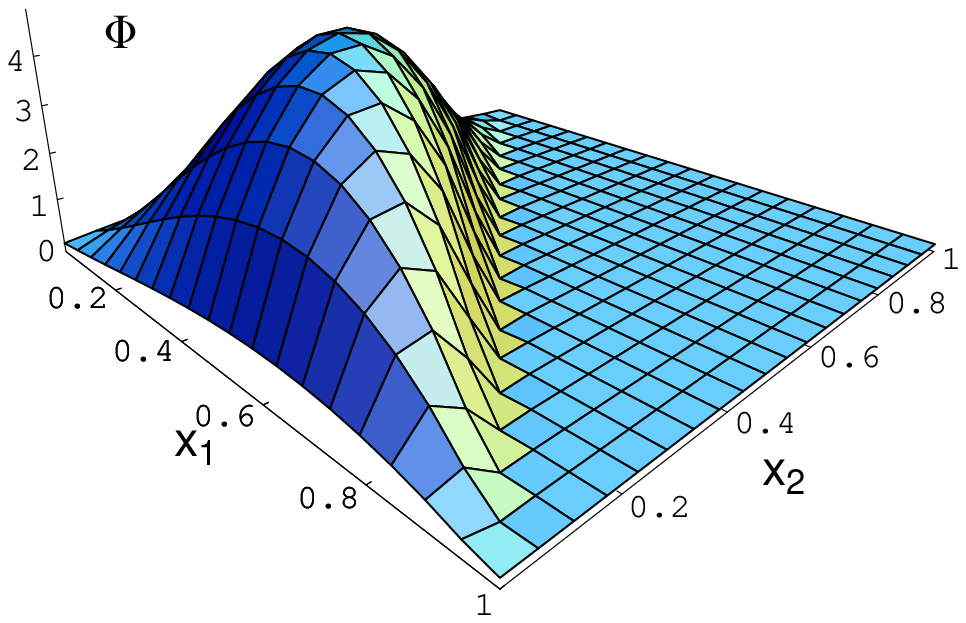,  width=13 pc}\\
} \caption{Comparison of the DA $\Phi$ between our model prediction
(left panel) and the data fit performed in
Ref.$^{16}$ (right panel).}
\end{figure}
In Table~\ref{Moments_comparison} we report our predictions for some moments
of the DA $\Phi$ in
comparison with the results from Refs. 
\refcite{DaPhi}, \refcite{KS}, \refcite{Dzi&Fran}, \refcite{Stef_Berg}, \refcite{Bolz_Kroll}, \refcite{LAT}.
 The moments are defined as
\begin{eqnarray}\label{momentsdef}
\Phi^{(l,m,n)} = \int_0^1 [dx]x_1^lx_2^mx_3^n\Phi(x_1,x_2,x_3)\bigg{/}\int_0^1 [dx]\Phi(x_1,x_2,x_3),\nonumber\\
\end{eqnarray}
where $[dx]=\delta(1 - \sum_{i=1}^3 x_i)\prod_{i=1}^3dx_i$.
\begin{table}
\tbl{Results for the moments of the DA $\Phi$ in our light-cone quark model in comparison with other model calculations: 
COZ~{\protect\cite{DaPhi}}, KS~{\protect\cite{KS}}, DF~{\protect\cite{Dzi&Fran}}, SB~{\protect\cite{Stef_Berg}},
data parameterization BK~{\protect\cite{Bolz_Kroll}} and lattice predictions LAT~{\protect\cite{LAT}}.
}
{\begin{tabular}{cccccccc}
\hline\hline
($l$,\,$m$,\,$n$) & $\, $ COZ$\, $ &$\, $ KS$\, $ & $\, $DF$\, $ & $\, $SB$\, $ & $\, $BK$\, $  & $\, $Our$\, $ & $\, $LAT$\, $ \\
\hline
$0\;\; 0\;\; 0\,$ & $\,    1    \,$  & $\,    1    \,$ & $\, 1   \,$ & $\, 1   \,$ & $\,  1  \,$  & $\,  1  \,$ & $\,  1  \,$\\
$1\;\; 0\;\; 0\,$ & $\,0.54\,$  & $\,0.46\,$ & $\,0.582\,$ & $\, 0.572 \,$ & $\,0.381\,$ & $\,0.346\,$  & $\,  0.394  \,$\\
$0\;\; 1\;\; 0\,$ & $\,0.18\,$  & $\,0.18\,$ & $\,0.213\,$ & $\, 0.184 \,$ & $\,0.309\,$ & $\,0.331\,$  & $\,  0.302  \,$\\
$0\;\; 0\;\; 1\,$ & $\,0.20\,$  & $\,0.22\,$ & $\,0.207\,$ & $\, 0.244 \,$ & $\,0.309\,$ & $\,0.323\,$ & $\,  0.304  \,$ \\
$2\;\; 0\;\; 0\,$ & $\,0.32\,$  & $\,0.27\,$ & $\,0.367\,$ & $\, 0.338 \,$ & $\,0.179\,$ &  $\,0.152\,$ & $\, 0.18  \,$\\
$0\;\; 2\;\; 0\,$ & $\,0.065\,$  & $\,0.08\,$ & $\,0.085\,$ & $\, 0.066 \,$ & $\,0.125\,$ & $\,0.142\,$ & $\, 0.132  \,$\\
$0\;\; 0\;\; 2\,$ & $\,0.09\,$  & $\,0.10\,$ & $\,0.83\,$ & $\, 0.170 \,$ & $\,0.125\,$ & $\,0.137\,$ & $\,  0.138  \,$\\
\hline\hline
\end{tabular}}
\label{Moments_comparison}
\end{table}
Our results are pretty close to the BK data
parameterization and to the very recent lattice predictions.\vspace{-4mm}
\section{Nucleon to pion TDAs}
In this section we apply the same formalism used for the DAs to the study of the TDAs in the baryonic sector and in particular for the description of the nucleon to pion transition. Such quantities can be analyzed in reactions such as backward pion electroproduction, e.g., $ep \rightarrow e'p'\pi_0$, and nucleon-antinucleon annihilation, e.g., 
$N\bar{N} \rightarrow \gamma^*\pi$.
As in Ref.~\refcite{Pireetal} we rewrite the general matrix element
describing a transition from a nucleon to a pion in terms of eight
TDAs: two vectorial $V_{1, 2}^{p\pi^0}(x_i,\xi,\Delta^2)$, two axial
$A_{1, 2}^{p\pi^0}(x_i,\xi,\Delta^2)$ and four tensorial
$T_{1,...,4}^{p\pi^0}(x_i,\xi,\Delta^2)$, \textit{i.e.}\vspace{-6mm}
\begin{eqnarray}\label{decomposition}
&&4\mathcal{F}\bigg(\langle\pi^0(p_\pi)|\epsilon^{ijk}u^i_\alpha(z_1n)
u^j_\beta(z_2n) d^k_\gamma(z_3n)|P(p_1,s_1)\rangle\bigg) = \nonumber\\
&&i\frac{f_N}{f_{\pi}}\Big[V_1^{p\pi^0}(\slashed{p}
C)_{\alpha\beta}(N^+)_\gamma + A_1^{p\pi^0}(\slashed{p}
\gamma^5C)_{\alpha\beta}(\gamma^5N^+)_\gamma\nonumber\\ && +
T_1^{p\pi^0}(\sigma_{p\mu}C)_{\alpha\beta}(\gamma^\mu N^+)_\gamma +
M^{-1}V_2^{p\pi^0}(\slashed{p}C)_{\alpha\beta}(\slashed{\Delta_T}N^+)_\gamma\nonumber\\
&&+
M^{-1}A_2^{p\pi^0}(\slashed{p}\gamma^5C)_{\alpha\beta}(\gamma^5\slashed{\Delta_T}N^+)_\gamma
+ M^{-1}T_2^{p\pi^0}(\sigma_{p\Delta_T}
C)_{\alpha\beta}(N^+)_\gamma\nonumber\\
&&+ M^{-1}T_3^{p\pi^0}(\sigma_{p\mu}C)_{\alpha\beta}(\sigma^{\mu\Delta_T}N^+)_\gamma +
M^{-2}T_4^{p\pi^0}(\sigma_{p\Delta_T}
C)_{\alpha\beta}(\slashed{\Delta_T}N^+)_\gamma\Big].\nonumber\\\vspace{-6mm}
\end{eqnarray}\vspace{-6mm}

Here, $ p_1 = (1 + \xi)p + \frac{M^2}{1 + \xi}n,$
$ p_\pi = (1 - \xi)p + \frac{m_\pi^2 + \Delta^2_T}{1 - \xi}n +
\Delta_T$, and $ \Delta = p_\pi - p_1 $, 
where we have defined $\xi = -\Delta^+/2P^+$ with $P =\frac{1}{2}(p_1
+ p_\pi)$. The TDAs depend on the light-cone momentum fractions $x_i$ ($\sum_i x_i = 2\xi$), 
on the skewedness parameter $\xi$ that describes the
change of longitudinal momentum from the initial and the final
state, and on $\Delta^2$.
As for the DAs, defining the left-hand side of
Eq.~(\ref{decomposition}) with the symbol
$D_{\alpha\beta,\gamma}^{\uparrow/\downarrow}$ it is possible to express the TDAs in
terms of eight of these matrix elements.
The calculations of the LCWF representation follows exactly the same line we used for the DAs. However to match the final one-pion state the nucleon state must now be composed at least by five partons with their dynamics encoded into two LCWFs, one for the $3q$ structure and one for $q\bar{q}$ pair, convoluted with a splitting function, $\phi_{\lambda' 0}^{\lambda(N, N\pi)}$, that weighs the different possible configurations.
For the details about the description of the nucleon as a bound state of a baryon surronded by its meson cloud we
refer to Ref.~\refcite{Barbara-MesonCloud}, while we postpone all the details regarding the calculation to a forthcoming paper\cite{NostroTDAs}.
We report here the general LCWF representation for the matrix element $D_{\alpha\beta,\gamma}^{\lambda}$,
\begin{eqnarray}\label{finalrepresentationD}
&&-\frac{24}{\sqrt{x_1x_2x_3}}\bigg(\frac{1}{2\xi}\bigg)^{\frac{3}{2}}
\sum_{\lambda_1, \lambda_2, \lambda_3}u_{+\alpha}(x_1P^+,
\lambda_1)u_{+\beta}(x_2P^+, \lambda_2)u_{+\gamma}(x_3P^+,
\lambda_3)\nonumber\\\nonumber&\times&\sum_{\lambda'}\int\mathrm{d}y\mathrm{d}^2\mathbf{k}_\perp\phi_{\lambda'
0}^{\lambda(N, N\pi)}(y, \mathbf{k}_{\perp})\sqrt{y(1 -
y)}\delta\bigg(1 - y -
\frac{p_\pi^+}{p_1^+}\bigg)
\nonumber\\&\times&
\int\frac{\prod_{i}\mathrm{d}^2\boldsymbol{k}_{i\perp}}{[2(2\pi)^3]^2}
\delta^{(2)}(\mathbf{p}_{\pi\perp} +
\mathbf{k}_{\perp})\delta^{(2)}\bigg(\sum_{i=1}^{3}\boldsymbol{k}_{i\perp}\bigg)\nonumber\\&\times&
\tilde{\Psi}_{\lambda'}^{[f]}\bigg(\bigg\{\frac{x_1}{2\xi},
\boldsymbol{k}_{1\perp}; \lambda_{1}, 1/2\bigg\}\bigg\{\frac{x_2}{2\xi},
\boldsymbol{k}_{2\perp}; \lambda_{2}, 1/2\bigg\}\bigg\{\frac{x_3}{2\xi},
\boldsymbol{k}_{3\perp}; \lambda_{3}, -1/2\bigg\}\bigg),
\end{eqnarray}
where, apart from some kinematical factors, it is easy to recognize the same
structure found for the DAs. 
In Eq.~(\ref{finalrepresentationD}) we note that the dependence from the light-cone momentum fraction in the LCWF is rescaled by a factor $2\xi$ with respect to the DAs case, as found in  
Ref.~\refcite{Pireetal} in the soft pion limit~\cite{Adler,Polyakov} ($\Delta_T=0,\xi\rightarrow1$) and for the
pion-nucleon GDAs in Ref.~\refcite{BraunGDAs}. In Fig.~2 the results for $V_1^{p\pi^0}$, $A_1^{p\pi^0}$ and $T_1^{p\pi^0}$
are shown.

\begin{figure}\begin{center}
\centerline{
\hspace*{1pc}
\epsfig{file=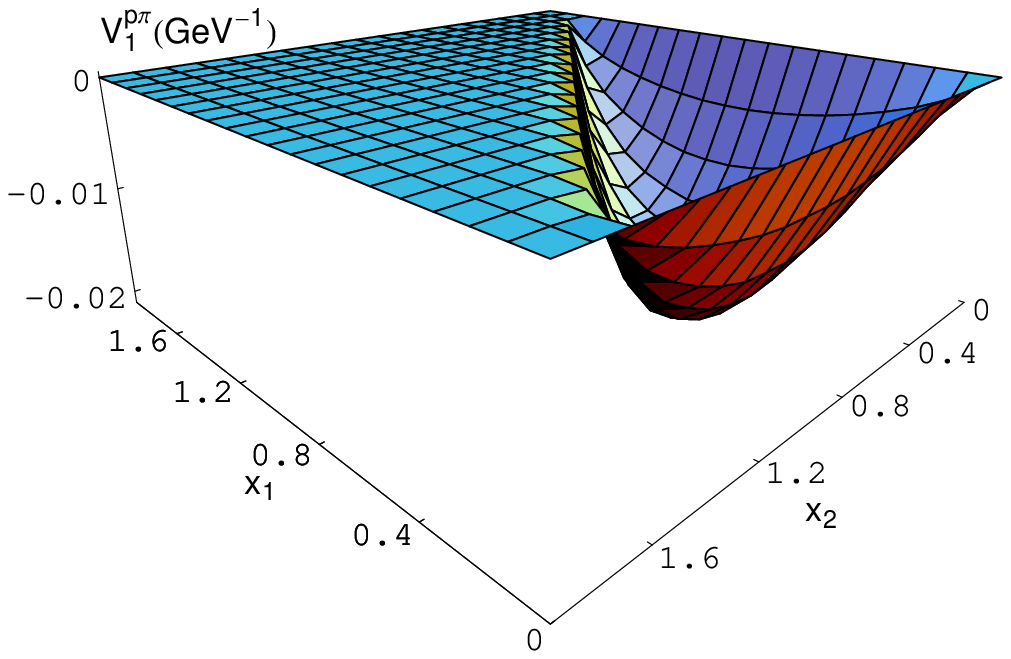,  width=14 pc}
\epsfig{file=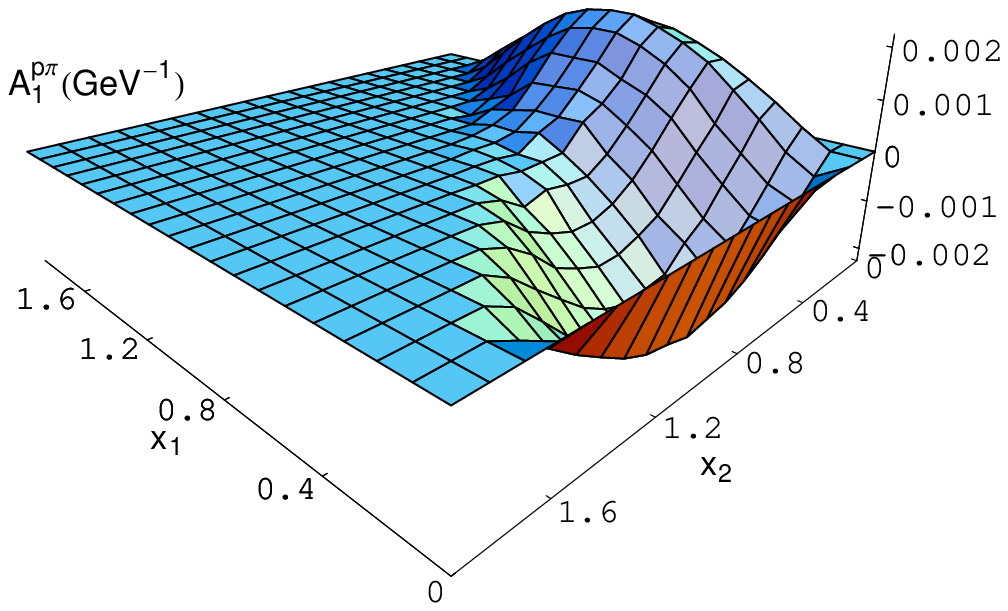,  width=14 pc}
}
\epsfig{file=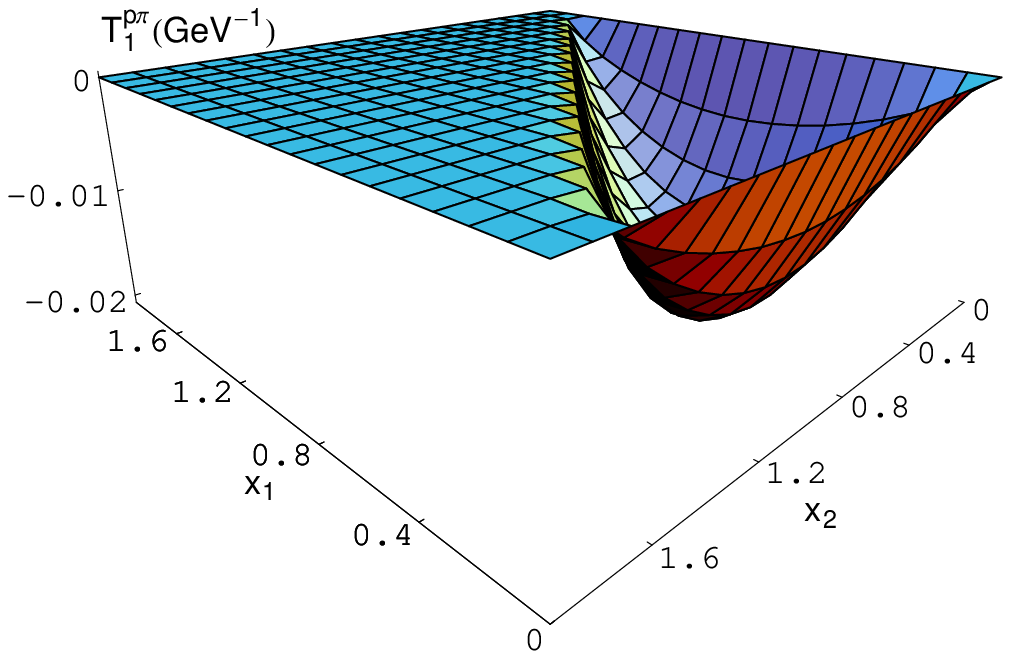,  width=14 pc}
\caption{Results for the TDAs $V_1^{p\pi^0}$, $A_1^{p\pi^0}$ and $T_1^{p\pi^0}$  at $\Delta^2$=-1GeV$^2$ and $\xi$=0.9.}
\end{center}\end{figure}

For a detailed numerical analysis of all the TDAs we refer to Ref.~\refcite{NostroTDAs}. 
\vspace{-2mm}
\section*{Acknowledgments}
M.P. acknowledges useful discussions with F.~Conti, A.~Courtoy and O.~Teryaev.
\vspace{-2mm}


\end{document}